\documentclass[final,a4paper]{IEEEtran}

\usepackage{graphicx}% Include figure files
\usepackage{dcolumn}% Align table columns on decimal point
\usepackage{bm}% bold math
\usepackage[english]{babel}
\usepackage{authblk}
\usepackage{amsmath}

\newcommand{\actualAlInP}{$\mbox{Al}_{0.59}\mbox{In}_{0.41}\mbox{P}$}
\newcommand{\InGaP}{$\mbox{In}_{0.52}\mbox{Ga}_{0.48}\mbox{P}$}

\newcommand{\InGaAsPxy}{$\mbox{In}_{1-x}\mbox{Ga}_{x}\mbox{As}_{y}\mbox{P}_{1-y}$}
\newcommand{\fig}[1]{\figurename{~(\ref{#1})}}
\newcommand{\eqn}[1]{Eq.~(\ref{#1})}

\begin{document}

\title{Investigation of Carrier Recombination Dynamics of InGaP/InGaAsP Multiple Quantum Wells for Solar Cells via Photoluminescence}% Force line breaks with \\
%\thanks{Footnote to title of article.}

\author{K. -H. Lee, K. W. J. Barnham, John S. Roberts, D. Alonso-\'Alvarez, N. P. Hylton, M. F\"uhrer, N. J. Ekins-Daukes
%\email{kan-hua.lee07@imperial.ac.uk}
%\homepage{http://www3.imperial.ac.uk/quanumphotovoltaics}
\thanks{K. -H. Lee was with Department of Physics, Imperial College London, London, United Kingdom. He is currently with Toyota Technological Institute, Nagoya, Japan. }
\thanks{K. W. J. Barnham, D. Alonso-
\'Alvarez, N. P. Hylton, M. F\"uhrer and N. J. Ekins-Daukes are with
Department of Physics, Imperial College London, London, SW7 2AZ, United Kingdom.}%\\This line break forced with 
%\textbackslash

\thanks{%
John S. Roberts is with EPSRC National Centre for III-V Technologies, Sheffield, S3 7HQ, United Kingdom.
}}%

\date{\today}% It is always \today, today,

%\pacs{quantum wells, 81.07.St,multijunction solar cells, 88.40.jp}% PACS, the Physics and Astronomy
\maketitle

\begin{abstract}
The carrier recombination dynamics of InGaP/InGaAsP quantum wells are reported for the first time. By studying the photoluminescence (PL) and time-resolved PL decay of InGaP/InGaAsP multiple-quantum-well(MQW) heterostructure samples, it is demonstrated that InGaP/InGaAsP MQWs have very low non-radiative recombination rate and high radiative efficiency compared to the control InGaP sample. Along with the analyses of PL emission spectrum and external quantum efficiencies, it suggests that this is due to small confinement potentials in the conduction band but high confinement potentials in the valence band. These results explain several features found in InGaP/InGaAsP MQW solar cells previously.
\end{abstract}

\section{Introduction}
\IEEEPARstart{A}{} novel InGaP/InGaAsP quantum well for tuning the absorption edge of the InGaP solar cell has been patented and implemented in commercial triple-junction solar cells\cite{Roberts:2011wp}\cite{Browne:2013ez}. Along with the GaAsP/InGaAs multiple-quantum-wells (MQWs) for GaAs solar cells, the absorption of both top cell and middle cell of standard InGaP/GaAs/Ge can both be fine tuned to maximize the energy yields. A simulation study has shown that the spectrally-tuned MQW 3J solar cells can improve the energy yields comparing to the standard InGaP/GaAs/Ge 3J solar cell with the same quality of bulk III-V materials\cite{Dobbin:2014gya}. This MQW is also a potential candidate for implementing the second junction of 1.99eV/1.51eV/1.15eV/Ge or 2.02eV/1.55eV/1.2eV/Ge quad-junction cells proposed in \cite{Toprasertpong:2015fu}.
%InGaAsP lattice-matched to GaAs has been used in semiconductor lasers as an aluminum-free option to generate laser at wavelengths from 730 nm to 808 nm as the active layer \cite{AlMuhanna:1998jra}\cite{Razeghi:1995ej}\cite{Yi:1995kfa}\cite{Fukunaga:1995jb}.
The main advantage of InGaAsP as the well material for top cell is that one can change both the compositions of group III and V elements simultaneously to tune the band gap while maintaining a fixed lattice constant. 
JDSU has demonstrated that triple-junction solar cells with dual-MQWs can achieve over 41\% efficiency at 500 suns\cite{Browne:2013ez}. Studies of this MQW structure were also reported in several publications\cite{Barnham:2010ev}\cite{Adams:2011dya}, mainly through external quantum efficiencies (EQE), I-V characteristics and detailed balance calculations. It was shown that this MQW has high external radiative efficiencies by measuring the photon coupling efficiency, however, the EQE of this MQW does not seem to be very efficient compared to GaAsP/InGaAs MQW solar cell. 

In order to resolve these issues, we performed detailed temperature-dependent photoluminescence experiments to investigate the carrier dynamics of the MQWs. We present a comparison between an InGaP/InGaAsP MQW and bulk InGaP control heterostructure sample. We further present the results from temperature dependent time-resolved PL decay and extract the effective recombination coefficients by fitting the experimental data with a rate model.

\section{Experiment and analysis}
Two undoped heterostructure samples were fabricated to study the material properties of InGaP/InGaAsP quantum wells. The sample structures are listed in Table~\ref{layerTable}. The first sample has InGaP/InGaAsP MQWs as the active layer, which has 20 repeats of 4.4-nm $\mbox{In}_{0.38}\mbox{Ga}_{0.62}\mbox{As}_{0.34}\mbox{P}_{0.66}$  wells separated by 13.1-nm \InGaP~barriers.
The thicknesses and compositions of the MQWs are comparable to earlier devices reported in \cite{Browne:2013ez}\cite{Lee:2012ee} and result in similar transition energies. The second one is a control sample, which is composed of bulk \InGaP~with the same active layer thickness as the former MQW sample. We will refer to them as MQW-HS and Control respectively in the rest of this paper.
Both samples were grown on GaAs substrates misorientated by 10 degrees towards the (001) plane to form disordered InGaP. The structures were grown using metalorganic chemical vapor deposition (MOCVD) and described in more detail in \cite{Roberts:2011wp}. The compositions of InGaAsP and InGaP are designed to be lattice-matched to GaAs.
Despite that having a control sample of bulk InGaAsP sample could be very beneficial to this study, we found it difficult to grow bulk InGaAsP of similar compositions to the well layers of MQW-HS\cite{Onabe:1982cda}.  Very recently this was achieved and reported in \cite{Jain:2016df} by using hydride vapor phase epitaxy.

The samples were excited by a PicoQuant diode pulsed laser with emission wavelength at 485 nm with a repetition rate of 80 MHz, and a pulse duration of a few hundreds picoseconds. The average power of the laser beam was 4 mW, and the beam focused onto the test sample was close to a Gaussian profile with $1/e^2$ radius of 49 $\mu \mbox{m}$. This gives a photon injection density at the Gaussian peak of $1.27\times 10^{12}~\mbox{cm}^{-2}$. 
%The peak injection density is equivalent to the order of $10^5$ suns. 
The PL spectrum was dispersed using a Princeton Instrument SP2500 monochromator and detected using a silicon photodetector through a Stanford Research SR530 lock-in amplifier.  The data was subsequently corrected to account for the spectral response of the detector and monochromator. 
\begin{table}
\centering
\caption{\label{layerTable}Layer structures of MQW-HS and Control. The nominal layer structure of the QWs are 13.1-nm InGaP barriers and 4.4-nm $\mbox{In}_{0.38}\mbox{Ga}_{0.62}\mbox{As}_{0.34}\mbox{P}_{0.66}$ wells.}
%\begin{ruledtabular}
\begin{tabular}{cccc}
layer name&MQW-HS(TS1111) & Control(TS1114) &thickness(nm)\\
\hline
cap&\InGaP & \InGaP & 5.8\\
cladding&\actualAlInP & \actualAlInP & 30\\
active layer&20xQWs & \InGaP & 348\\
%active&\InGaP & \InGaP & 37.8\\
cladding&\actualAlInP & \actualAlInP & 45\\
substrate&GaAs&GaAs& \\
\end{tabular}
\footnotetext{}
%\end{ruledtabular}
\end{table}

\subsection{Steady-state PL}
\begin{figure}
	\centering
	\includegraphics[width=\columnwidth]{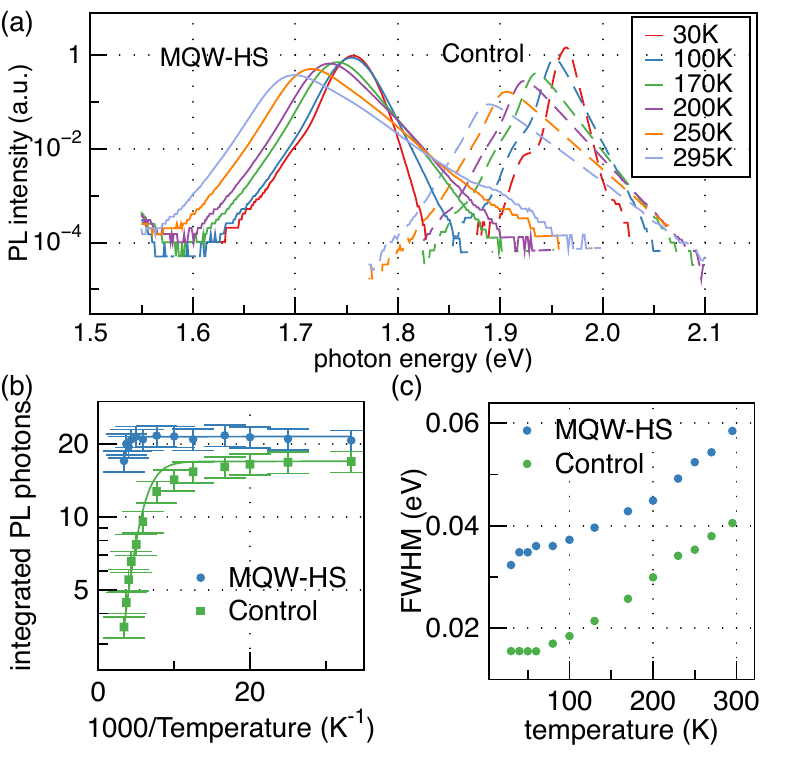}
	\caption{The temperature dependent PL spectra of MQW-HS and Control. The PL spectra of MQW-HS are plotted in solid lines, whereas the PL spectra of Control are plotted in broken lines. All the PL spectra are normalized by the same factors. (b) and (c) are rescaled spectrally integrated PL and the full-width half maximum of the PL, respectively. The solid lines in (b) are fittings using equation (2) in \cite{Lambkin:1990gya}. The fitted activation energies of MQW-HS and Control are 0.16 eV and 0.065 eV, respectively.}
	\label{fig:TS1111_TS1114_PL}
\end{figure}

The normalized PL spectra of MQW-HS and Control measured at temperatures between 30 K and 295 K are plotted in \figurename~\ref{fig:TS1111_TS1114_PL}(a). The peak PL photon energy of MQW-HS is 1.69 eV at room temperature which is comparable to the absorption edge reported in \cite{Lee:2012ee}. The peak PL photon energy of Control is around 1.9 eV at room temperature, which is a characteristic of disordered InGaP. We note that ordered InGaP typically shows room temperature PL around 1.82 eV \cite{Yeo:1997ux}\cite{Dabkowski:1988ctb}\cite{Capaz:1993kj}.

\figurename~\ref{fig:TS1111_TS1114_PL}(b) shows the spectrally integrated, normalized PL. The spectrally integrated PL of MQW-HS saturates at temperatures below 200 K, suggesting that non-radiative recombination process do not play a dominant role in this temperature range. Also, a comparison of the absolute magnitude of the PL emission from the two samples shows that their integrated PL intensities are similar at 30 K. Thus we infer that the recombination in both samples is likely to be highly radiative at low temperatures. The radiative efficiency at room temperature, which is defined as the ratio between the radiative recombination and the overall recombination of electron-hole pairs, can be estimated as the ratio of the spectrally integrated PL at room temperature and low temperature. By this measure, the radiative efficiency of MQW-HS at room temperature is found to be 84\%, while the radiative efficiency of Control is only 20\%. The radiative efficiency of MQW-HS is comparable to the values reported in \cite{Lee:2012ee}, which was obtained from measurements of the photon coupling efficiencies of dual-MQW dual-junction devices. In \cite{Lee:2012ee}, the highest measured photon coupling efficiency of the MQW InGaP cell is 52\%, which results in a radiative efficiency of 68\%. The temperature dependent PL was fitted using the rate model described in \cite{Lambkin:1990gya}. The fitted activation energy of MQW-HS and Control are 0.16 eV and 0.065 eV, respectively.
The measured radiative efficiency of MQW-HS is higher than the values measured in \cite{Lee:2012ee} because the instantaneous injection intensity is higher and there are no emitter or base layers which cause additional recombination losses, i.e. here we measure the internal quantum efficiency of the MQWs only, not a complete device. \figurename~\ref{fig:TS1111_TS1114_PL}(c) presents a plot of the PL full width at half maximum height, showing that the emission of MQW-HS is very broad. This will be discussed further in the next section.

\subsection{TRPL decay}
\begin{figure}
	\centering
		\includegraphics[width=\columnwidth]{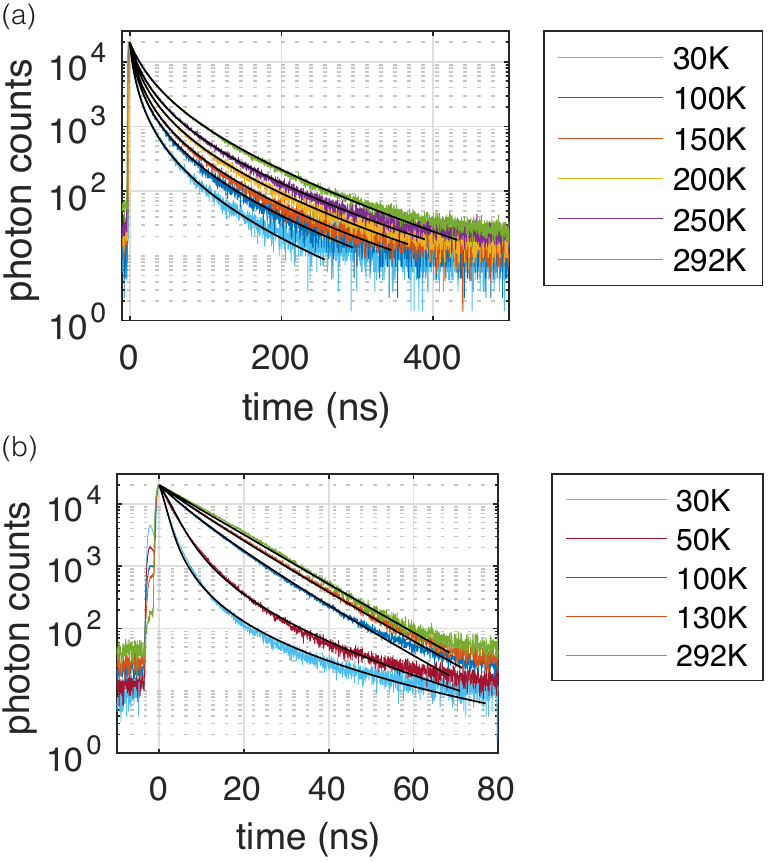}
	\caption{Measured and fitted TRPL decays of (a)MQW-HS and (b)Control at various temperatures.} 
	\label{fig:Figures_TRPLPlot}
\end{figure}

TRPL decay curves at the PL peak wavelengths were measured by time-correlated single photon counting to further study the recombination mechanisms of these samples. TRPL decay of MQW-HS and Control were illuminated at the same pulse power and beam size described in the previous section, but the repetition rate is reduced to a few MHz in order to measure longer decay times. The measured PL decays are plotted in \figurename~\ref{fig:Figures_TRPLPlot}.

The PL decay is determined by the instantaneous excess carrier density, which can be described by the rate equation\cite{Smith:2002ef}:
\begin{equation}
	\label{mainRateEq2}
	\frac{\partial \sigma_n}{\partial t}
	=-A\sigma_n-\frac{B}{w}\sigma_n^2-\frac{C}{w^2}\sigma_n^3+D\nabla^2\sigma_n
\end{equation}

The linear term coefficient $A$ is the function of Shockley-Read-Hall recombination, surface recombination and monomolecular recombination, $B$ is bimolecular radiative recombination coefficient, $C$ is the coefficient that relates to Auger recombination and $D$ is the diffusion coefficient of the carriers, and $\sigma_n$ is the excess sheet carrier density\cite{Smith:2002ef}\cite{ahrenkiel1993_full}.  
This assumes that the excited carriers are distributed uniformly within an effective layer of thickness $w$, and the recombination coefficients $A$, $B$ and $C$ are constant throughout the PL decay. Although separating the contributions of the real recombination coefficients and photon re-absorption factors is difficult, it is the effective recombination rates, $A$, $B$ and $C$, that are of the most interest for devices since they are directly linked to the diffusion length of the carriers. Denoting $B_s\equiv{} B/w$ and $C_s\equiv{}C/w^2$, the recombination coefficients $A_s$, $B_s$ and $C_s$ can be retrieved by solving the rate equation to fit the TRPL decay data. 
In this way, we can reduce the three-dimensional rate equation to a two-dimensional one without affecting $A$ because $A=A_s$. The approximations of $w$ in each sample and their impacts on resulting $B$ will be discussed later.
  
The diffusion coefficient of lowly-doped InGaAsP/InP QWs is reported to be of the order of few $ \mbox{cm}^2\mbox{sec}^{-1}$ at room temperatures\cite{Smith:2002ef}\cite{Marshall:2000te}. We thus use this to approximate $D$ in our samples. The effect of diffusion at this level is found to be negligible in the decay time scale considered here. The difference of fitted $A_s$ due to varying $D$ from zero to $10~\mbox{cm}^2\mbox{sec}^{-1}$
is around $10^5~\mbox{sec}^{-1}$, and the difference of fitted $B_s$ caused by varying $D$ in the same range is negligible.

In most cases, setting the coefficient of the third-order term $C$ to be zero yields reasonable fit, suggesting that the Auger recombination is negligible. 

The effective layer of thickness $w$ of MQW-HS and Control need to be considered respectively due to different carrier redistribution mechanisms in bulk and MQWs. Since both InGaP and InGaAsP have the absorption coefficients around $1.5\times10^5~\mbox{cm}^{-1}$ at the laser wavelength\cite{Goldberg:WeMcBXFD}\cite{Goldberg:xZDu1wia}, 64\%($\sim 1-1/e$) of the carriers are absorbed in the first 75 nm of the active layer of both samples, so this value can be regarded as a lower limit of $w$ for both samples. For MQW-HS, most of the carriers are captured and recombine in the quantum well, showing that the carriers are transferred from one well to another by thermal escape or photon re-absorption. Therefore, $w$ of MQW-HS has a range from few wells to the total thickness of 20 wells depending on how uniform that the carriers are distributed. For Control, the excess carriers are redistributed by diffusion. At room temperature, the diffusion length of the carriers is $\sim1700$ nm, based upon a diffusion coefficient of $1~\mbox{cm}^2\mbox{sec}^{-1}$ \cite{Goldberg:xZDu1wia} and carrier lifetime ($1/A_s$) of 30 ns, so we can assume that the excess carriers in Control are redistributed within the whole bulk InGaP layer. As a result, the upper bound of $w$ of Control can be estimated as the whole active layer thickness. At lowest measured temperature (30K), the diffusion coefficient is of the order of $0.1~\mbox{cm}^2\mbox{sec}^{-1}$ according to Einstein's relation and the mean lifetime is around 5 ns based on an observation of \figurename~\ref{fig:Figures_TRPLPlot}(b), the lower bound of $w$ of Control is thus around 200 nm. As a result, $w$ of Control covers a large portion of the active layer even in low temperatures.

With the approximations and assumptions mentioned above, (\ref{mainRateEq2}) can be solved using the laser profile as the initial condition of $\sigma_n(r)$. 
The total PL decay can then be calculated from the solution $\sigma_n(r,t)$ using the relation\cite{Ahrenkiel:1993je}:
\begin{equation}
	\label{integratePhotonsEq}
	\phi \propto B_s\int_S\sigma_n(r)[(\sigma_n(r)+\sigma_D)]dS
\end{equation}
where $\sigma_D$ is the sheet background doping density of the samples and $S$ is the front surface of the sample. The background doping concentration $N_D$ is of the order of $\sim 10^{15}~\mbox{cm}^{-3}$. The sheet background doping density can thus be written as $\sigma_D=N_D w$, which gives a sheet background doping density $\sigma_D$ an upper bound $\sim 10^9~\mbox{cm}^{-2}$  by taking $w$ as the thickness of the whole active layers. Since the excess density range covered by this TRPL decay measurement is from $10^{9}$ to $10^{12}~\mbox{cm}^{-2}$, we have $\sigma_n(r) \gg \sigma_D$. (\ref{integratePhotonsEq}) thus becomes
\begin{equation}
	\phi \propto B_s\int_A \sigma_n^2(r)dA
\end{equation}

\begin{figure}
	\centering
		\includegraphics[width=\columnwidth]{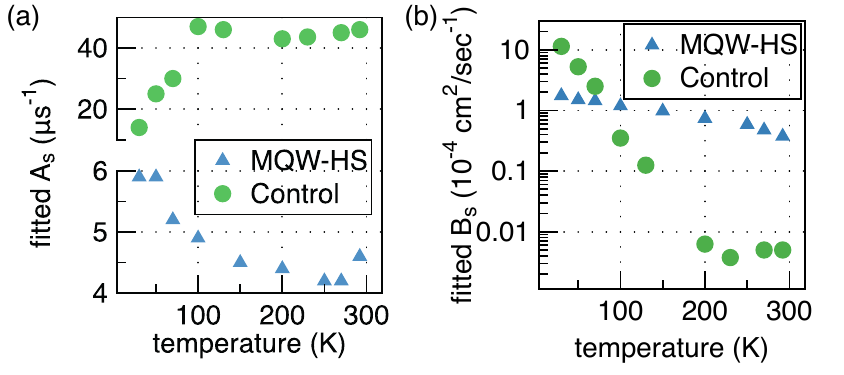}
	\caption{The fitted values of $A_s$ and $B_s$ of MQW-HS and Control in (\ref{mainRateEq2}).}
	\label{fig:Figures_TRPLFittingPlot}
\end{figure}

%Qualitative feature
The fitted TRPL decay and the fitted values are plotted in \figurename~\ref{fig:Figures_TRPLPlot} and \figurename~\ref{fig:Figures_TRPLFittingPlot}.
This model fits the PL decay well at most timescales. However, the long lived tails of the experimental PL decay are not fitted very well, which may be due to the occupancies of long-lived trap states in the samples that slow the non-radiative recombination rates at late stage of the decay.

The fitted $B_s$ of MQW-HS is of the order of $10^{-4}~\mbox{cm}^2/\mbox{sec}$, thus the radiative recombination coefficient $B$ is between $10^{-11}$ to $10^{-10}~\mbox{cm}^3/\mbox{s}$, depending on the exact value of $w$. The range of $B$ is comparable to the radiative recombination coefficients of InP/InGaAsP MQWs reported in \cite{Smith:2002ef}, \cite{Fox:1989ha} and \cite{Wintner:1984ij}. From quantum mechanical calculations and experiment results on $B$ reported in \cite{Garbuzov:1982kk} and \cite{ahrenkiel1993_B}, we expect $B$ decreases as the temperature increases. Also, $w$ would increase with temperature because the carrier diffusion and thermal escapes in quantum wells are enhanced at higher temperatures. Since $B_s=B/w$, the trends of the fitted $B_s$ of both MQW-HS and Control presented in \figurename~\ref{fig:Figures_TRPLFittingPlot} matches the expected temperature dependent behaviors of $B$ and $w$.

The fitted $A_s$ of MQW-HS and Control have different trends. Below 100 K, the fitted $A_s$ of Control increases with temperature, and then stays at the same level above 100 K. This result suggests that $A_s$ of Control is dominated by non-radiative recombination, which generally increases with temperature. However, the fitted $A_s$ of MQW-HS decreases with temperature, which suggests that radiative recombination is still dominant at high temperatures. This matches the qualitative behavior of the TRPL decays shown in \figurename~\ref{fig:Figures_TRPLPlot}. When the carrier injection density $n$ is low or the $A$ coefficient is high ($A \gg Bn$), the PL decay is dominated by the term $An$ and therefore is a straight line in the logarithmic plot. In contrast, if the carrier density $n$ is high or $A$ is low ($A \ll Bn$), the decay will not be single-exponential and the instantaneous decay rate will change with the remaining excess carrier densities. Therefore, at high carrier density, the initial decay of the TRPL is dominated by radiative recombination and dependent on $B$ and $n$, and the decay rate in the late stage is determined by $A$. At low temperatures, both MQW-HS and Control have non-single-exponential PL decay. However, at high temperatures, Control is dominated by single-exponential decay but MQW-HS is still non-single-exponential, suggesting that MQW-HS has a much lower $A$ coefficient.

The $A_s$ coefficient of MQW-HS is around $10^6~\mbox{sec}^{-1}$, which is one order of magnitude lower than Control at all temperatures. Since $A_s$ is the upper limit of the non-radiative recombination rates, this supports the observation of the steady-state PL that the quantum well sample has much higher radiative efficiency than the bulk sample.

In \cite{Smith:2002ef}, the $A_s$ of undoped InP/InGaAsP MQWs at room temperature is reported to be $2\times 10^5~\mbox{sec}^{-1}$, which is one order of magnitude lower than the fitted $A_s$ of MQW-HS. This suggests that the quality of InGaP/InGaAsP MQWs is not as high as InP/InGaAsP MQWs.
 
\section{Discussion}
\label{sec:discussion}

The steady-state PL measurement has shown that the radiative efficiency of the InGaP/InGaAsP MQWs is high, and this is consistent with the observed trend of $A_s$ against temperature in the TRPL decay experiment. In addition, the fitted recombination rate $A_s$ of the MQW sample is much lower than the control sample. Also, the FWHM of this MQW sample's PL emission is broader than the PL emission of the control sample.

\begin{figure}
	\centering
	\includegraphics[width=\columnwidth]{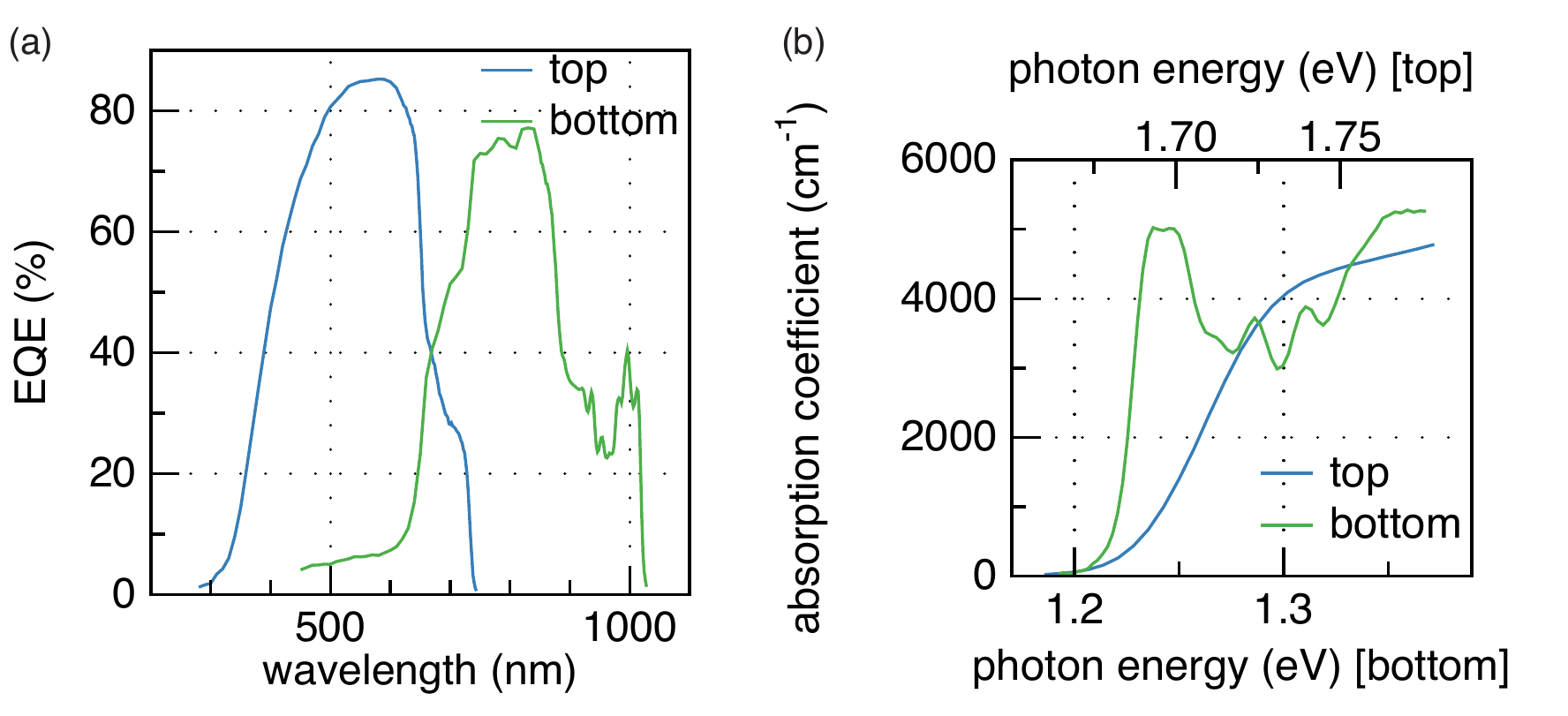}
	\caption{The (a) measured EQE and (b) extracted absorption coefficients of the subcells in the dual-MQW dual-junction device described in \cite{Lee:2013dy}.}
\label{fig:2698EQE}
\end{figure}

The broad PL emission of MQW sample is consistent with the modest onset of the absorption near the band-edge evident from the external EQE of this MQW InGaP cell. \figurename~\ref{fig:2698EQE} shows the EQE and absorption coefficients of a dual-MQW dual-junction solar cell reported in \cite{Lee:2013dy}. This cell has a top cell that has MQWs with same design as MQW-HS and a GaAs bottom cell with GaAsP/InGaAs MQWs. The absorption coefficients are calculated from EQEs by assuming a infinite diffusion length, i.e., 
\begin{equation}
\alpha(E)=-\log(1-\mbox{EQE}(E))/L 
\end{equation}
where $\alpha(E)$ is the absorption coefficient, $E$ is the photon energy, and $L$ is the thicknesses of the total MQWs layer. We can observe that the absorption edge of the bottom cell's MQW is much sharper than the top cell's MQW. This suggests a tail of states in the QW or type II transitions\cite{Bastard:1988vq}. It is worth noted that this characteristic of EQE can also be found in this type of MQW cell fabricated by other facilities, such as the ones reported in \cite{Browne:2013ez}.

This feature can be explained by an approximation of band alignment using electron affinities. The electron affinity of InGaAsP can be approximated by interpolation \cite{Adachi:1982ifb}:
\begin{equation}
\begin{split}
Q(x,y)&=(1-x)yB_{\mbox{InAs}}+xyB_{\mbox{GaAs}} \\
&+(1-x)(1-y)B_{\mbox{InP}}+x(1-y)B_{\mbox{GaP}}
\end{split}
\label{eqn:InGaAsP_interp}
\end{equation}
where $B_{InAs}$, $B_{GaAs}$, $B_{InP}$ and $B_{GaP}$ are the parameters of InAs, GaAs, InP, and GaP listed in \tablename{~\ref{table:InGaAsPmat}}.

\begin{figure}
	\centering
	\includegraphics[width=\columnwidth]{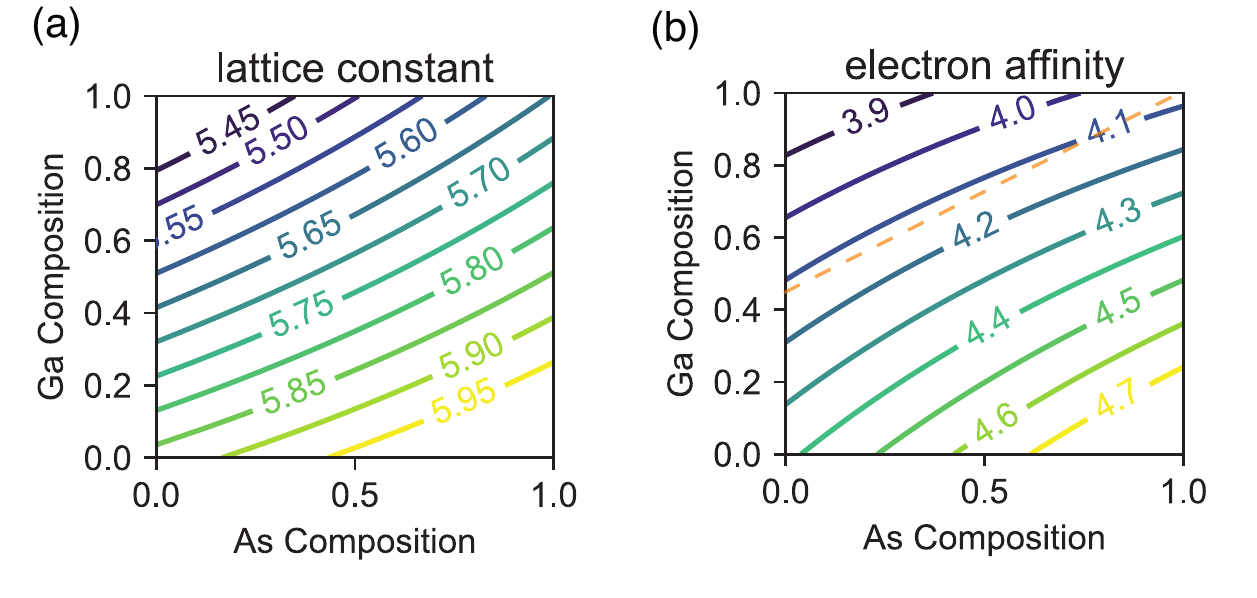}
	\caption{The calculated (a) lattice constant($\mbox{A}^\circ$) and 
	(b) electron affinity (eV) contours of \InGaAsPxy~ as a function of $x$ and $y$ using 
	\eqn{eqn:InGaAsP_interp} and the parameters in Table~{\ref{table:InGaAsPmat}}. The orange dashed line in (b) represent the \InGaAsPxy ~lattice-matched to GaAs.}
	\label{fig:InGaAsPParametersMap}
\end{figure}

\fig{fig:InGaAsPParametersMap} shows the band gaps and lattice constants of \InGaAsPxy~ as the function of Ga and As compositions respectively. The dashed line in \fig{fig:InGaAsPParametersMap}(b) represent the combination of $x$ and $y$ that keeps its lattice constant equal to GaAs (5.64~$\mbox{A}^{\circ}$). As shown in the figure, this line almost overlaps with the contour of 4.1 eV , which is very close to the electron affinity of GaAs and InGaP. This suggests that the confinement potential of electrons in the quantum well is small, which causes the slow onset of absorption as shown in \figurename~\ref{fig:2698EQE}.

This argument can also explain the fitted activation energy of MQW-HS (0.16 eV). This value is close to the difference between the peak PL wavelengths of MQW-HS and Control. Since the result of TRPL shows that the MQW is very radiative, the fitted activation energy should correspond to the energy barrier for carrier escape from the well to the barrier. As a result, either the confinement potential in the conduction band or valence band is large whereas the other one is small. The estimation of electron affinities suggest that it is the confinement potentials in the valence band that accounts for this activation energy. The shallow confinement potential in conduction band may also explain the high leakage current found in semiconductor lasers with InGaAsP active layer lattice-matched to GaAs\cite{Yi:1995kfa}. 

\begin{table}
\centering
\caption{The material parameters of the binary materials for estimating the band gaps, electron affinities and lattice constants of InGaAsP\cite{Adachi:1982ifb}.}
\begin{tabular}{cccc}
 & band gap (eV)  & lattice constants ($\mbox{A}^\circ$)  & electron affinity (eV) \\ 
\hline
GaP & 2.776 & 5.4512 & 3.8 \\ 
GaAs & 1.415 & 5.6533 & 4.07 \\ 
InP & 1.35 & 5.8688 &  4.38\\ 
InAs & 0.36 & 6.0584 & 4.9 \\ 
\end{tabular}
\label{table:InGaAsPmat}
\end{table}

\section{Conclusions}
The carrier recombination dynamics of InGaP/InGaAsP multiple quantum wells is presented for the first time. The spectrally integrated PL shows that InGaP/InGaAsP MQWs can achieve 84\% radiative efficiency at high excitation, which is comparable to the estimations obtained from fitting the dark currents and direct measurement of the radiative coupling in \cite{Barnham:2010ev} and \cite{Lee:2012ee} respectively.
%max photon coupling eta is around 52%
In addition, the temperature dependent recombination coefficients show that the MQWs at same excitation levels as PL experiment is radiative dominant across the whole temperature range, while the bulk InGaP sample is dominated by non-radiative recombination. Along with the observations of PL spectrum and spectrally integrated PL, this result indicates that non-radiative recombination in the MQWs is suppressed by the quantum confinement potentials in the wells. The estimated electron affinity suggests that strong quantum confinement potentials mainly come from the valence band, which is consistent to the activation energy of non-radiative recombination obtained by temperature-dependent PL.

\section*{Acknowledgments}
This work is partly supported by QuantaSol Ltd. and the European Commission within the FP7 Research Framework Program NGCPV.

\bibliographystyle{IEEEtran}
\bibliography{mqwtopcellv2,support_book}% Produces the bibliography via BibTeX.

\end{document}